\documentclass[12pt]{amsart}
\usepackage{amsmath}
\usepackage{amssymb}
\usepackage{graphicx}
\usepackage[utf8]{inputenc}
\usepackage[T1]{fontenc}
\newtheorem{lemma}{Lemma}

\newcommand{\tr}{{\rm Tr }}

\newcommand{\tG}{\Gamma}

\newcommand{\bra}{\langle}
\newcommand{\ket}{\rangle}
\newcommand{\vp}{\varphi}

\newcommand{\C}{\mathbb{C}}
\newcommand{\N}{\mathbb{N}}

\newcommand{\R}{\mathbb{R}}
\newcommand{\be}{\begin{equation}}
\newcommand{\eeq}{\end{equation}}
\newcommand{\bet}{\begin{equation*}}
\newcommand{\eeqt}{\end{equation*}}
\newcommand{\bea}{\begin{eqnarray}}
\newcommand{\eeqa}{\end{eqnarray}}
\newcommand{\beat}{\begin{eqnarray*}}
\newcommand{\eeqat}{\end{eqnarray*}}

\newcommand{\h}[1]{\mathcal{#1}}
\newcommand{\hil}{\mathcal{H}}

\newcommand{\hA}{\mathcal{A}}
\newcommand{\hB}{\mathcal{B}}

\newcommand{\hK}{\mathcal{K}}

\newcommand{\br}{\mathcal{B}(\R)}
\newcommand{\lh}{L(\mathcal{H})}

\newcommand{\sfq}{\mathsf{Q}}
\newcommand{\sfp}{\mathsf{P}}

\begin{document}
\title{Position and momentum tomography}

\author{Jukka Kiukas}
\address{Institute for Mathematical Physics, Technical University of Braunschweig, Braunschweig, Germany}
\email{jukka.kiukas@utu.fi}
\author{Pekka Lahti}
\address{Department of Physics and Astronomy, University of Turku, Turku, Finland}
\email{pekka.lahti@utu.fi}
\author{Jussi Schultz}
\address{Department of Physics and Astronomy, University of Turku, Turku, Finland}
\email{jussi.schultz@utu.fi}

\begin{abstract}
We illustrate the use of the statistical method of moments for determining the position and momentum distributions of a quantum object from the statistics of a single measurement. The method is used for three different, though related, models; the sequential measurement model, the Arthurs-Kelly model and the eight-port homodyne detection model. In each case, the method of moments gives the position and momentum distribution for a large class of initial states, the relevant condition being the exponential boundedness of the distributions. 
\end{abstract}

\maketitle

\section{Introduction}
One of the main problems of quantum mechanics deals with
 the possibility of measuring together the position and momentum distributions $\rho^\sfq$
and $\rho^\sfp$ of a quantum system prepared in a state $\rho$. 
The basic structures of quantum mechanics dictate that there is no (joint) measurement 
which would directly give both the position and momentum distributions and that, for instance, any determination of the position
distribution $\rho^\sfq$ necessarily disturbs the  system such that the initial momentum distribution $\rho^\sfp$ 
gets drastically changed.

In recent years two important steps have been taken in solving this problem. 
First of all, the original ideas of
Heisenberg \cite{Heisenberg1927} have finally  been brought to a successful end with the seminal paper of Werner \cite{WernerIII} which
gives operationally feasible necessary and sufficient conditions for a 
measurement to serve as an approximate joint measurement of the position and momentum distributions, including also the
inaccuracy-disturbance aspect of the problem. The second breakthrough in studying this question comes from a reconstruction of the state $\rho$ from 
a single informationally complete measurement, notably realized optically by an eight-port homydyne detection \cite{LeonhardtI}, \cite[p.147-155]{LeonhardtII} (for a rigorous quantum mechanical treatment, see \cite{JukkaVII}). In conjunction with an explicit state reconstruction formula (known at least for the Husimi-distribution \cite{Dariano1994}), this allows one to immediately determine the distributions of any given observables.

If one is only interested in determining the position and momentum distributions $\rho^\sfq$ and $\rho^\sfp$, 
it is obviously unnecessary to reconstruct the entire state; one should be able to do this with less information. 
Here we will use the statistical method of moments to achieve a scheme for position and momentum tomography, i.e. the reconstruction of the position and momentum distributions from the measured statistics. 
The price for using moments is, of course, that they do not exist for all states, and even when they do, they typically do not determine the distribution uniquely. 
Hence, we restrict here to the states for which the position and momentum distributions are exponentially bounded. 
We note that this is an operational condition and can, in principle, be tested for a given moment sequence \cite{LPY}.

We consider three different, though related, measurement schemes based on the von Neumann model, Sect.~\ref{vNmodel}, 
and the balanced homodyne detection technic,
Sect.~\ref{homodyne}.
The first model is a sequential measurement of a standard position measurement of the von Neumann type \cite{vN} followed by any momentum
measurement, Sect.~\ref{sequential}.
The second (Sect.~\ref{AKB}) builds on the  Arthurs-Kelly model \cite{AK65} as developed further
by Busch \cite{Busch1982} whereas the third (Sect.~\ref{homodyne}) model uses the quantum optical realizations of position and momentum as
the corresponding quadrature observables of a (single mode) signal field implemented by balanced homodyne detection \cite{JukkaVI}.
In Sect.~\ref{simultaneous} we apply the method of moments to determine both the position and momentum distributions $\rho^\sfq$ and $\rho^\sfp$
from the actually measured statistics.
Finally, we compare our method with
the state reconstuction method, Sect.~\ref{info}. There we also comment briefly the possibility of inverting convolutions. 
We begin, however, with quoting the basic no-go results on the position-momentum joint/sequential measurements.

\section{No joint measurements}
There are many formulations of the basic fact that position and momentum of a quantum object cannot be  measured jointly,
or, equivalently, that, say, any position measurement `destroys' all the information on the momentum prior to the measurement.
In this section we recall one of the most striking formulations of this fact. To do that we fix first some  notations.

Let $\hil$ be a complex separable Hilbert space and  $\lh$ the set
of bounded operators on $\hil$. Let $\Omega$ be a nonempty set and
$\hA$ a $\sigma$-algebra of subsets of $\Omega$. The set function
$E:\hA\to\lh$ is a {\em semispectral measure}, or normalized positive
operator measure, POM, for short, if the set function $\hA\ni
X\mapsto \bra\psi|E(X)\psi\ket\in\C$ is a probability measure for
each $\psi\in\hil_1$, the set of unit vectors of $\hil$. 
We denote this probability measure by $p^E_\psi$.
A semispectral measure $E$ is  a
spectral measure if it is projection valued, that is, all the
operators $E(X), X\in\hA$, are projections. If $\hil$ is the Hilbert
space of a quantum system, then the observables of the system are
represented by  semispectral measures $E$ and the numbers
$\bra\psi|E(X)\psi\ket$, $X\in\hA$,  $\psi\in \hil_1$,
are the measurement outcome probabilities for $E$ in a vector state
$\psi$. 
An observable is called sharp if it is represented by a
spectral measure. Otherwise, we call it unsharp.
 Here we
consider only the cases where the measurement outcomes are real
numbers, that is, $(\Omega,\hA)$ is the real Borel space $(\R,\br)$,
or, pairs of real numbers, in which case $(\Omega,\hA)$ is
$(\R^2,\hB(\R^2))$.
The position and momentum distributions $\rho^\sfq$ and $\rho^\sfp$ are just the probability measures $p^\sfq_\rho$ and $p^\sfp_\rho$
defined by $\sfq$ and $\sfp$  together with a density matrix (mixed state) $\rho$.

An observable $M:\h B(\R^2)\to L(\hil)$ has two marginal observables $M_1$ and $M_2$ defined by
the conditions 
$M_1(X)=M(X\times\R)$ and $M_2(Y)=M(\R\times Y)$ for all $X,Y\in\br$. 
Any measurement of $M$ constitutes a joint measurement of $M_1$ and $M_2$. On the other hand, any two 
observables $E_1$ and $E_2$ admit a joint measurement (or equivalently a sequential joint measurement) if there is
an observable (on the product value space) $M:\h B(\R^2)\to L(\hil)$ such that $E_1=M_1$ and $E_2=M_2$.
The following result is crucial:\footnote{This result seems to be well-known, and part of the proof goes back to Ludwig \cite[Theorem 1.3.1]{Ludwig}. However, we were unable to identify a full proof in the literature, and so we give one in the appendix}

\begin{lemma}\label{apu3} Let $M:\h B(\R^2)\to L(\hil)$ be a semispectral measure, such that one of the marginals is a spectral measure. Then, for any $X,Y\in \h B(\R)$, 
$M_1(X)M_2(Y)=M_2(Y)M_1(X)$, that is, the marginals commute with each other,  and
$M(X\times Y) = M_1(X)M_2(Y)$, that is, $M$ is of the product form.
\end{lemma}

Assume that $M:\h B(\R^2)\to L(\hil)$ is an observable with, say, the first marginal observable $M_1$ being the
position of the object. Then
$M_1$ and $M_2$ commute with each other, and due to the maximality of the position observable $\sfq$
any $M_2(Y)$ is a function of $\sfq$. 
Therefore, $M_2$ cannot represent (any nontrivial version of) the momentum observable. Similarly,
if one of the marginal observables is the momentum observable, then the two marginal observables are pairwisely
commutative, and the effects of the other marginal observable are functions of the momentum observable.

\section{Position/momentum measurements}

It is a basic result of the quantum theory of measurement that each observable (sharp or unharp) admits a realization
in terms of a measurement scheme, that is, each observable has a measurement dilation \cite{Ozawa}. 
In particular, this is true for the position and momentum observables $\sfq$ and $\sfp$. However,
due to the continuity of these observables they do not admit any repeatable measurements \cite{Ozawa, Luzack}. 
In fact, the known realistic models
for position and momentum measurements serve only as their approximative measurements which constitute
$\sfq$ and $\sfp$ -measurements   only in some 
appropriate limits. Here we consider two such models, the standard von Neumann model and the optical
version of a $\sfq$, resp. $\sfp$, -measurement in terms of a balanced homodyne detection. 
Before entering these models we briefly recall the notion of intrinsic noise of an observable 
and the corresponding characterization of noiseless measurements.

For an observable  $E:\br\to\lh$ the $k^{\rm th}$ moment operator is the (weakly defined) symmetric operator $E[k]=\int_\R x^k\,dE$ with its
natural (maximal) domain $D(E[k])$. 
In particular, the number $\langle\psi|E[k]\psi\rangle=\int_\R x^k\,dp^E_\psi$ is the $k^{\rm th}$ moment of the probability measure $p^E_\psi$.
The (intrinsic) noise of $E$ is defined as $N(E)=E[2]-E[1]^2$, and it is known to be
positive, that is, $\langle \vp | N(E)\vp\rangle \geq 0$ for all $\vp\in D(E[2])\cap D(E[1]^2)$.
If the first moment operator $E[1]$ of $E$ is selfadjoint, then $E$ is sharp exactly when $E$ is noiseless, that is,
$N(E)=0$ \cite{JukkaIV}.\footnote{The selfadjointness of the first moment operator is crucial for this condition. 
Indeed, if, for instance,  one restricts the spectral measure of the momentum observable  $\sfp$  in $L^2(\R)$ by a projection $\sfq(I), I=[a,b]$, 
to get a POM $\tilde\sfp:Y\mapsto \sfq(I)\sfp(Y)\sfq(I)|_{L^2(I)}$
acting on $L^2(I)$,  one has $\tilde\sfp[k]=\tilde\sfp[1]^k$ for all $k$, and thus also $N(\tilde\sfp)=0$, though the first moment $\tilde\sfp[1]$  is 
only a densely
defined symmetric operator \cite{DKP}. This is also an example of the variance free observables as discussed in \cite{Werner1990}. 
A noiseless observable is variance free, but due to the domain conditions the reverse implication may not be true.}
We recall also that the first moment operator $E[1]$ of an observable alone is never sufficient to determine the
actual observable. In statistical terms, the first moment information (expectation) $\langle\psi|E[1]\psi\rangle$, $\psi\in\hil_1$, does not suffice to
determine the measured observable $E$.

\subsection{The von Neumann model}\label{vNmodel}
Consider the von Neumann model of a position measurement of an object confined to move in one spatial dimension \cite[Sect. VI.3]{vN}, 
see also e.g. \cite[Sect. II.3.4]{OQP}. 
Let $\hil=L^2(\R)$ be the Hilbert space
of the object system, and let $Q$ denote its position operator. 
We let $\sfq$ denote the spectral measure of $Q$.
To measure $Q$ we couple it with the momentum $P_0$ of the probe
system, with the Hilbert space $\hK=L^2(\R)$, and we monitor the shifts in probe's position $Q_0$, with the spectral measure $\sfq_0$. 
Let $U=e^{-i\lambda Q\otimes P_0}$ be the 
unitary measurement coupling, with a coupling constant $\lambda >0$, 
$\phi\in\hK$, $\parallel\phi\parallel=1$, the initial probe state,
and let $V_\phi:\hil\to\hil\otimes\hK$ denote the embedding $V_\phi(\vp)=\vp\otimes\phi$. 
The actually measured observable of
the object system is then given by measurement dilation formula
$$
E(X)=V_\phi^*U^*I\otimes \sfq_0(X)UV_\phi, \quad X\in\br.
$$
A direct computation shows that $E$ is an unsharp position, with the effects
\begin{equation}\label{sumeapaikka}
E(X)= (\chi_X\ast f)(Q),
\end{equation}
where $\chi_X\ast f$ denotes the convolution of the characteristic function $\chi_X$ of the set $X\in\br$ with the probability density
$f(x)=\lambda |\phi(-\lambda x)|^2$.

\subsubsection{Limiting observable}
The actually measured observable $E$ depends on two parameters: the coupling constant $\lambda$ and the initial probe state  $\phi$,
 that is, $E=E^{\lambda,\phi}$. The structure of the effects (\ref{sumeapaikka})  suggests that the semispectral measure $E$ comes close to 
the spectral measure $\sfq$ whenever the convolution $\chi_X\ast f$ comes close to $\chi_X$. This evident fact can be quantified in various ways.

Due to the convolution structure of $E$, the geometric distance between the observables $E$ and $\sfq$
can easily  be computed \cite{WernerIII}, and one finds that
$$
d(E,\sfq)=\frac{1}{\lambda} \int \vert x\vert \vert \phi(x )\vert^{2} dx,
$$
showing that whenever the integral is finite, the geometric distance tends to zero as $\lambda$ increases, or $\vert \phi (x)\vert^{2}$ becomes more 
sharply concentrated around the origin. It follows from the definition of the geometric distance, that $d(E,\sfq )=0$ implies $E=\sfq$. 
However, this does not settle the question of the limit $E\to\sfq$ in either of the two possible intuitive meanings.
For that we use the method of moments.

In order to be able to determine the moment operators of the unsharp position observable $E$, we assume that
$\phi\in C^\infty_\downarrow(\R)$, so that, in particular $\phi\in D(Q_0^k)$ for each $k\in\N$.
In that case the moment operators $E[k]$ can all be computed,\footnote{Some of the technical details behinds these computations have been studied in
\cite{KLY08}.} 
and they turn out to be polynomials of  degree $k$ of $Q$,
that is, $D(E[k])=D(Q^k)$, and
\begin{equation}\label{moments}
E[k]=\sum_{i=0}^k\binom{k}{i}\lambda^{-i}\langle{\phi|Q_0^i\phi}\rangle\,Q^{k-i}.
\end{equation}
Therefore, in particular, 
on $D(E[2])=D(Q^2)$, one has 
$N(E)=\frac 1{\lambda^2}{\rm Var}\,(Q_0,\phi)I$,
suggesting, again, that, for a fixed $\phi$, if $\lambda$ is large, then the noise $N(E)$ is small, or, for a fixed $\lambda$,
if  ${\rm Var}\,(Q_0,\phi)$ is small, then, again, $N(E)$ would be small.
But, again, the precise meaning 
of the limit $E\to \sfq$ in either of the cases
$\lambda\to\infty$ or ${\rm Var}\,(Q_0,\phi)\to 0$  waits to be qualified. 

Consider first the limit $\lambda\to\infty$, so that, the operator measures are actually $E^\lambda$, with the moment operators 
$E^\lambda[k]$ of (\ref{moments}). Let $D$ be the linear hull of the Hermite functions, so that 
$D \subset D(Q^k)=D(E^\lambda[k])$ for all $k$ (and for all $\lambda$),
and 
\begin{equation}\label{datalimit}
\lim_{\lambda\to\infty}\langle\psi|E^\lambda[k]\psi\rangle = \langle\psi|Q^k\psi\rangle
\end{equation}
for all $\psi\in D$ and $k\in\N$. 
Due to the exponential boundedness of the Hermite functions, the moments
 $\langle\psi|Q^k\psi\rangle$, $k\in\N$, 
of the probability measure  $p^Q_\psi$ determine it uniquely \cite{Freud}.
Since $D$ is a dense subspace, the probability measures $p^Q_\psi$, $\psi\in D$, determine, by polarization, the spectral
measure $\sfq$ of $Q$.
To conclude that on the basis of the statistical data (\ref{datalimit}), the observable
 $E^\lambda$  would converge to  $\sfq$, one needs to know that also  $E^\lambda$ is determined by
its moment operators  $E^\lambda[k]$, $k\in\N$,  on  $D$. 
Again,  for all  $\psi\in D$, the probability measures $p^{E^\lambda}_\psi$ are exponentially bounded, so that  each $p^{E^\lambda}_\psi$
is determined by its moments $p^{E^\lambda}_\psi[k]=\langle\psi|E^\lambda[k]\psi\rangle$, $k\in\N$. 
Hence, by polarization, $E^\lambda$ is determined by the
numbers $\langle\psi|E^\lambda[k]\psi\rangle$, $k\in\N,\psi\in D$.

Let now $\lambda_n$, $n\in\N$, be an increasing sequence of the coupling constants, with $\lambda_n \to\infty$, and let $(E^n)_{n\in\N}$
be the sequence of the semispectral measures $E^{\lambda_n}$. The above results show that $\sfq$ is the moment limit of the sequence
$(E^n)_{n\in\N}$ on $D$, that is, we may write
\begin{equation}\label{limit}
\lim_{n\to\infty} E^{n}=\sfq
\end{equation}
(on $D$ in the sense of moment operators),
for further technical details, see \cite{JukkaVI}).
We remark that in this case also the effects $E^n(X)$ tend weakly to the projections $\sfq(X)$ for all $X\in\br$ whose boundaries
$\overline X\cap\overline{X'}$ are of Lebesgue measure zero, \cite{JukkaVI}.

The corresponding limits for the case ${\rm Var}\,(Q_0,\phi)\to 0$ can similarly been worked out, for instance, if $\phi$ is chosen to be the
Gaussian state $\phi_n(x)= \left(\frac{n}{\pi}\right)^{1/4} e^{-nx^2/2}$, and one considers the limit $n\to\infty$.

\subsubsection{Indirectly measured observable}
In addition to obtaining the limit (\ref{limit}),
formula (\ref{moments}) can also be solved directly  for the numbers $\langle\psi|Q^k\psi\rangle$, $\psi\in D, \in \N$. Indeed, 
one may write recursively
\begin{equation}\label{recursion}
\bra \psi|Q^k\psi\ket = \langle\psi|E[k]\psi \rangle -\sum_{i=1}^{k}\binom ki \lambda^{-i}\langle\phi|Q_0^i\phi\rangle  \bra \psi|Q^{k-i}\psi\ket  , \ \ k\in\N.
\end{equation}
These numbers
 are the moments of the probability distributions $\rho^Q$ for $\rho=|\psi\rangle\langle\psi|$.  
Due to the exponential boundedness of these distributions they are
uniquely determined by their moments $\langle\psi|Q^k\psi\rangle$, $k\in\N$, and  by the density of $D$, the polarization identity then implies that this 
statistics is sufficient to determine also the position observable $\sfq$. Though the actually measured observable in this model is the unsharp
position $E^{\lambda,\phi}$, the measurement statistics allows one to determine 
also directly, without any limit considerations,
the `unobserved' sharp position $\sfq$.
In Sect.~\ref{info} we discuss still another method to obtain the position distribution $\rho^\sfq$, $\rho=|\psi\rangle\langle\psi|$,  from the actually
measured distribution $f*\rho^Q$ by inverting the convolution.

\subsection{The balanced homodyne detection observable}\label{homodyne}

The balanced homodyne detection scheme is a basic measurement scheme
in many quantum optical applications, including continuous variable
quantum tomography as well as continuous variable quantum
teleportation. Such a measurement scheme determines an observable
$E^z$ which depends on the coherent state $|z\rangle$, $z\in\C$, of
the auxiliary field. An important property of these observables is
that on the level of statistical expectation values they agree with
the quadrature observables $Q_\theta=\frac 1{\sqrt 2}(e^{-i\theta}a+
e^{i\theta}a^*)$, $z=re^{i\theta}$, of the relevant field mode, with
the annihilation operator $a$. 
The explicit structure of these
observables $E^z$ has been studied in great detail and, in
particular, their moment operators are determined
\cite{JukkaVI}.

To express the relevant results here, we let $D(a)$ stand for the
domain of the annihilation operator (which, in terms of the fixed
number basis $\{|n\rangle\}_{n\in\N}\subset\hil$, is
$D(a)=\{\vp\in\hil\,|\, \sum_{n\in\N}n\,|\langle n|\vp\rangle|^2<\infty\}$), and $N=a^*a$ is the corresponding (selfadjoint) number operator.
 The first and the
second moment operators of such a balanced homodyne detection
observable $E^z$, $z=re^{i\theta}$,  are known to  be as follows:
\begin{eqnarray*}
E^z[1]|_{D(a)}    
&=&     
Q_\theta|_{D(a)},\\
E^z[2]|_{D(a^2)}  
&=& 
(Q_\theta|_{D(a)})^2+ \frac 12
r^{-2}N.
\end{eqnarray*}
Here e.g.  $E^z[k]|_{D(a^k)}$    
denotes the restriction of the $k^{\rm th}$ moment operator $E^z[k]$
of $E^z$ to the domain $D(a^k)$, $k=1,2$.
By definition, the noise operator $N(E^z)$ has the domain $D(N(E^z))= D(E[2])\cap D(E[1]^2)$,
which includes the set $D(N)=D(a^2)$ because of the above operator relations. Hence,
$\frac 12 r^{-2} N\subset N(E^z)$. But $N(E^z)$ is symmetric and $N$ selfadjoint, so that
$N(E^z)= \frac 12 r^{-2} N$.  This would again suggests that in the limit $r=|z|\to\infty$, the intrinsic noise $N(E^z)$
goes to zero and thus the measured observable would approach the quadrature observable $\sfq_\theta$. 
Like in the previous case, Sect.~\ref{vNmodel}, this limit requires  further considerations.

Actually,
the restrictions of all the moment  operator $E^z[k]$ on the domains $D(a^k)$, $k\in\N$, can be determined, and they are of the form
\begin{equation}\label{hdmoments}
E^z[k] |_{D(a^k) }= (Q_\theta |_{D(a^k)})^k+\frac 1{r^2} C_k(r,\theta),
\end{equation}
where $C_k(r,\theta)= \sum_{{n,m }_{\, n+m\leq k}}c^k_{n,m}(r,\theta)(a^*)^na^m$, and each $c^k_{n,m}$ is a bounded complex function on
$[1,\infty)\times[0,2\pi)$ \cite{JukkaVI}.
Let $D_{coh} =\ {\rm lin}\{|w\rangle\,|\, w\in \C\}$, so that $D_{coh}$  is a dense subspace contained in all  $D(a^k)$, $k\in\N$.
 For each unit vector $\psi\in  D_{coh}$, the probability measure $p_\psi^{E^z}$ is exponentially bounded so that it is determined by its
moment sequence $\langle\psi|E^z[k]\psi\rangle, k\in\N$. 
Since $D_{coh}$ is dense, these probability measures define again the whole operator measure $E^z$ \cite{JukkaVI}.

Let now $(r_n)$ be a sequence of positive numbers converging to infinity. For this choice, let $z_n(\theta)=r_ne^{i\theta}$,
where the phase $\theta\in[0,2\pi)$ is also fixed, and let $E^n$ be the corresponding balanced homodyne detection observable.
By the above results it now follows that the spectral measure $\sfq_\theta$ is the only 
moment limit of  the sequence of observables $(E^n)$. Moreover, for any unit vector $\psi$, $\lim_{n\to\infty}p^{E^n}_\psi(X)=
p^{\sfq_\theta}_\psi(X)$ for all $X\in\br$ whose boundary $\overline{X}\cap\overline{X'}$ is of Lebesgue measure zero \cite{JukkaVI}.
In this sense one can say that the high amplitude limit of the balanced homodyne detection scheme serves as  an experimental implementation
of a quadrature observable.

Again, one may solve the statistical moments $\langle\psi\,|\,Q_\theta^k\psi\rangle$ from (\ref{hdmoments}) for all $\psi\in D(a^k)$.
However, in this case they are not directly expressible in terms of actually measured moments $\langle\psi\,|\,E^z[k]\psi\rangle$. 
The high amplitude limit is needed for that end.

To close this section we mention that in a recent paper Man'ko {\em et al} \cite{Manko} has proposed to use the first and second moments of the 
measurement statistics of the (limiting) balanced homodyne detection observables  associated with the phases $\theta, \theta+\frac \pi{2},\, \theta+\frac\pi{4}$,
to empirically test the uncertainty relations for the conjugate quadratures (associated with $\theta, \theta+\frac \pi{2})$. 
Clearly, for any $\psi\in D(a^2)$, with the choice $\theta =0$ and notations $Q_0=Q, Q_{\frac\pi{2}}=P$,
\begin{eqnarray*}
{\rm Var}_\psi(E^r){\rm Var}_\psi(E^{ir}) &=& 
\left(\langle E^r[2] \rangle - \langle E^r[1] \rangle^2  \right)  \left( \langle E^{ir}[2] \rangle   -  \langle E^{ir}[1] \rangle^2  \right) \\
&=& 
\left(\langle Q^2\rangle + \textstyle{\frac 12}\,r^{-2}\langle N \rangle  -\langle Q \rangle^2 \right)
\left(\langle P^2\rangle + \textstyle{\frac 12}\,r^{-2}\langle N \rangle  -\langle P \rangle^2 \right)
\\
&=&\left({\rm Var}_\psi(Q) + \textstyle{\frac 12}\,r^{-2}\langle N \rangle \right)
\left({\rm Var}_\psi(P) + \textstyle{\frac 12}\,r^{-2}\langle N \rangle \right) \geq \frac 14,
\end{eqnarray*}
which allows one to test the statistics in this respect for any $|z|=r$. The marginal statistics of the limiting eight-port homodyne detection
observables of Section~\ref{homodyne} leads to a similar inequality, except with the lower bound 1. We wish to point out that the test proposed in \cite{Manko} is actually an experimental check for the correctness of the quantum mechanical description of balanced homodyne detection, since any violation of the above inequality would suggest that the description is incorrect.

\section{Combining position and momentum measurements}

We shall go on to combine the above measurement schemes to produce sequential and joint measurements for position and momentum.
We consider first the sequential application of a standard position measurement with any momentum measurement. Sections~\ref{AKB} and \ref{homodyne}
deal with the Arthurs-Kelly model and the eight-port homodyne detection scheme.

\subsection{Sequential combination}\label{sequential}
Consider an approximate position measurement, described by the von Neumann model, followed by a sharp momentum measurement. 
This defines a unique sequential joint observable, a covariant phase space observable $G^{\lambda,\phi}:\mathcal{B}(\R^{2})\rightarrow L(\hil)$, with the marginals
\begin{eqnarray*}
G^{\lambda,\phi}_{1}(X) &=&(\chi_{X} \ast e )(Q),\\
G^{\lambda,\phi}_{2} (Y) &=& (\chi_{Y} \ast f)(P).
\end{eqnarray*}
Here we have the probability densities $e(q) =\lambda \vert \phi (-\lambda q)\vert^{2}$ and $f(p) =\frac{1}{\lambda}\vert \hat{\phi}(-\frac{p}{\lambda} )\vert^{2}$, 
where $\phi\in\hil$, $\Vert\phi\Vert =1$, is the initial probe state, and $\hat{\phi}$ denotes the Fourier transform of $\phi$. If $\phi\in C^{\infty}_{\downarrow}(\R)$, 
we have $\phi\in D(Q^{k}_0)\cap D(P^{k}_0)$ for each $k\in\N$, in which case the moment operators of the marginal observables are
\begin{eqnarray}
 G^{\lambda,\phi}_{1}[k] &=& \sum^{k}_{i=0} \binom{k}{i} \lambda^{-i} \langle \phi \vert Q^{i}_0\phi\rangle Q^{k-i}, \label{marginaalit11}\\
 G^{\lambda,\phi}_{2} [k]&=&\sum^{k}_{i=0} \binom{k}{i} \lambda^{i} \langle\phi\vert P^{i}_0\phi\rangle P^{k-i}. \label{marginaalit12}
\end{eqnarray}
As shown before, we have 
$$
\lim_{\lambda\rightarrow\infty} \langle\psi\vert G^{\lambda,\phi}_{1}[k]\psi\rangle =\langle\psi \vert Q^{k}\psi\rangle
$$
for all $\psi\in C^{\infty}_{\downarrow}(\R)$. In the case of the second marginal we see that for any $\psi\in C^{\infty}_{\downarrow}(\R)$ there are values of $k\in\N$ 
for which $\langle\psi \vert G^{\lambda,\phi}_{2}[k]\psi\rangle $ tends to infinity as $\lambda$ increases. 
For example, the limit of the second moment is never finite
since $\langle \phi \vert P^{2}_0\phi\rangle$ is always non-zero. 
That is, the limits of the moments of the probability measure 
$X\mapsto \langle\psi \vert G^{\lambda,\phi}_{2}(X)\psi\rangle  =\langle\psi\vert G^{\lambda,\phi}(\R\times X)\psi\rangle$ are not moments of any determinate probability measure, and hence they do not determine any observable.

Another way to look at the limits of the marginal observables is to choose a sequence of initial probe states $(\phi_{n})_{n\in\N}\subset L^{2}(\R)$, such that $\vert\phi_{n} \vert^{2}$ approaches the delta distribution as $n$ increases. For example, choose the Gaussian states
$$
\phi_{n} (x) =\left (\frac{n}{\pi} \right)^{1/4} e^{-n\frac{x^2}{2}},
$$
in which case the explicit forms of the moment operators $ G^{\lambda,n}_{1}[k]$ and $G^{\lambda,n}_{2}[k]$ can easily be computed:
\begin{eqnarray}
 G^{\lambda,n}_{1}[k] &=& \sum^{k}_{i=0,\ i\textrm{ even}}\binom{k}{i} \frac{\lambda^{-i}}{\sqrt{n^{i}\pi}} \tG\left(\frac{i+1}{2} \right) Q^{k-i}, \label{marginaalit21}\\
G^{\lambda,n}_{2}[k] &=& \sum^{k}_{i=0,\ i\textrm{ even}}\binom{k}{i} \lambda^{i}\sqrt{\frac{n^{i}}{\pi}} \tG\left(\frac{i+1}{2} \right) P^{k-i},\label{marginaalit22}
\end{eqnarray}
where $\tG$ denotes the gamma function. Taking the limit $n\rightarrow\infty$ one gets a result similar to the one considered before ($\lambda\rightarrow\infty$).

As expected, the limit procedures cannot give both the $\sfq$  and $\sfp$ -distributions, but as it is obvious from (\ref{marginaalit11}-\ref{marginaalit12}) 
and (\ref{marginaalit21}-\ref{marginaalit22})
the method of moments can again be used. We return to that in Sect.~\ref{simultaneous}.

Again, the convolution structure allows one to easily compute the distances between the marginals and the sharp position and momentum observables. One finds that
\begin{eqnarray*}
d(G^{\lambda,\phi}_1 ,\sfq) &=& \frac{1}{\lambda} \int \vert x\vert \vert \phi (x)\vert^2 dx,\\
d(G^{\lambda,\phi}_2 ,\sfp ) &=& \lambda \int \vert x\vert \vert \hat{\phi}(x)\vert^2 dx,
\end{eqnarray*}
showing, that the product of the distances does not depend on $\lambda$. 
Since the distances are Fourier-related, their product has a positive lower bound, that is, 
$\inf_{\phi\in\hil_1} d(G^{\lambda,\phi}_1 ,\sfq)\cdot d(G^{\lambda,\phi}_2 ,\sfp )>0$. 
For example, in the case of the Gaussian initial states $\phi_n$ one has $d(G^{\lambda,n}_1,\sfq )\cdot d(G^{\lambda,n}_2,\sfp) =\frac{1}{\pi}$ for all $n\in\N$.

\subsection{Arthurs-Kelly model}\label{AKB}
The Arthurs-Kelly model \cite{AK65} as developed further by  Busch \cite{Busch1982} (see also \cite{Raymer, TSJ}) is based on the von Neumann model of an approximate measurement. It consists of standard position 
and momentum measurements performed simultaneously on the object system. Consider a measuring apparatus consisting of two probe systems, 
with associated Hilbert spaces $\hil_1$ and $\hil_2$. Let $\phi_{1}\otimes\phi_{2}\in\hil_1 \otimes \hil_2$ be the initial state of the apparatus. 
The apparatus is coupled to the object system, originally in the state $\varphi\in\hil$, by means of the coupling
\begin{equation}\label{coupling}
U= e^{-i\lambda Q\otimes P_{1} \otimes I_{2}} e^{i\mu P\otimes I_{1} \otimes Q_{2} },
\end{equation}
which changes the initial state of the object-apparatus system $\Psi_{0} =\psi\otimes \phi_{1} \otimes \phi_{2}$ into $\Psi =U\Psi_{0}$. The final state $\Psi$ has the position representation 
$$
\Psi (x,y,z) = \psi (x+\mu z) \phi_{1} (y-\lambda x) \phi_{2} (z).
$$
Notice, that the coupling (\ref{coupling}) is a slightly simplified version of the one used by Arthurs and Kelly.
However, it does not change any of our conclusions.

The measured covariant phase space observable $G$ is determined from the condition
$$
\langle \psi\vert G(X\times Y)\psi\rangle =\langle \Psi \vert I\otimes \sfq (\lambda X) \otimes \sfp (\mu Y)\Psi\rangle,
$$
for all $X,Y\in\br$, and the marginal observables $G_{1}$ and $G_{2}$ turn out to be
\begin{eqnarray}
G_{1}(X) &=& (\chi_{X} \ast (e_{0}\ast \vert \phi^{(\mu)}_{2}\vert^2 ))(Q),\label{AKmarginaali1}\\
G_{2}(Y) &=& (\chi_{Y} \ast (f_{0}\ast \vert \hat{\phi}^{(\lambda)}_{1}\vert^2 ))(P)\label{AKmarginaali2} ,
\end{eqnarray}
where $e_{0}$ and $f_{0}$ are the probability distributions related to the original single measurements, 
i.e. $e_{0}(q) =\lambda \vert \phi_{1}(-\lambda q)\vert^{2}$ and $f_{0}(p) =\mu \vert \hat{\phi}_{2} (-\mu p )\vert^{2}$, 
and we have used the scaled functions 
$\phi_{1}^{(\lambda)} (q) =\sqrt{\lambda} \phi_1(\lambda q)$ and $\phi_{2}^{(\mu)} (p) =\frac{1}{\sqrt{\mu}} \phi_{2} (\frac{p}{\mu})$ . 
If we choose the initial state of the apparatus to be such that $\phi_{1},\phi_{2} \in C^{\infty}_{\downarrow}(\R)$, the moment operators can be computed:
\begin{eqnarray}
 G_{1}[k] &=& \sum_{n=0}^{k}\sum_{i=0}^{n} \binom{k}{n} \binom{n}{i} \lambda^{-(n-i)}(-\mu)^{i}\langle \phi_{1}\vert Q^{n-i}_{1}
 \phi_{1}\rangle \langle\phi_{2}\vert Q^{i}_{2}\phi_{2}\rangle Q^{k-n},\label{marginaalit31}\\
G_{2}[k] &=& \sum_{n=0}^{k}\sum_{i=0}^{n} \binom{k}{n} \binom{n}{i} \mu^{-(n-i)}(-\lambda)^{i}\langle \phi_{2}\vert P^{n-i}_{2} 
\phi_{2}\rangle \langle\phi_{1}\vert P^{i}_{1}\phi_{1}\rangle P^{k-n}.\label{marginaalit32}
\end{eqnarray}

It is clear from equations (\ref{AKmarginaali1} -\ref{AKmarginaali2}),  that the $\sfq$- and $\sfp$-distributions 
cannot  be simultaneously obtained as limits of the marginals, since the distributions $e_{0}\ast \vert \phi^{(\mu)}_{2}\vert^2$ and $f_{0}\ast \vert \hat{\phi}^{(\lambda)}_{1}\vert^2$ 
cannot both be arbitrarily sharply concentrated. However, equations (\ref{marginaalit31}-\ref{marginaalit32}) show that the method of moments can  be used.

\subsection{Eight-port homodyne detector}\label{homodyne}
\begin{figure}
\includegraphics[width=12cm]{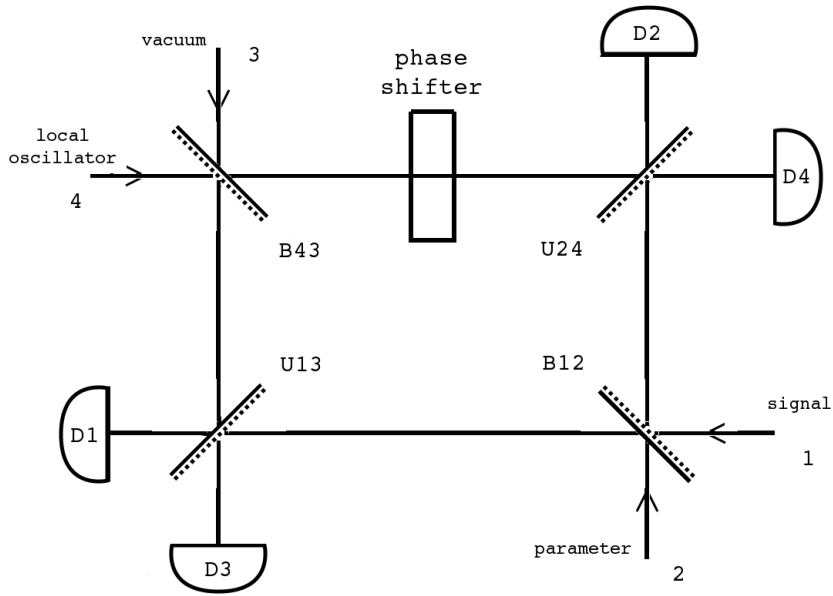}
\caption{The eight-port homodyne detector}\label{detector}
\end{figure}

The eight-port homodyne detector \cite{LeonhardtI,LeonhardtII} consists of the setup shown in Figure \ref{detector}. 
The detector involves four modes and the associated Hilbert spaces will be denoted by $\hil_1$, $\hil_2$, $\hil_3$ and $\hil_4$. 
Mode 1 corresponds to the signal field, the input state for mode 2 serves as a parameter which determines the observable to be measured, 
and mode 4 is the reference beam in a coherent state. The input for mode 3 is left empty, corresponding to the vacuum state. 
We fix a photon number basis $\{ \vert n\rangle \vert n\in\N \}$ for each $\hil_{j}$, so that the annihilation operators $a_{j}$, 
as well as the quadratures $Q_{j} =\frac{1}{\sqrt{2}} (\overline{a_{j}^{*} +a_{j}})$, $P_{j} =\frac{i}{\sqrt{2}} (\overline{a_{j}^{*} -a_{j}})$, 
and the photon number operators $N_{j} =a_{j}^{*} a_{j}$ are defined for each mode $j=1,2,3,4$.

The photon detectors $D_{j}$ are considered to be ideal, so that each detector measures the sharp photon number $N_{j}$. 
The phase shifter is represented by the unitary operator $e^{i\xi N_{4}} $, where $\xi$ is the shift. There are four 50-50-beam splitters $B_{12}$, $B_{43}$, $U_{13}$, $U_{24}$, each of which is defined by its acting in the coordinate representation: 
\begin{equation}\label{split}
L^{2}(\R^{2} ) \ni \Psi \mapsto \big( (x_{i},x_{j }) \mapsto \Psi (\frac{1}{\sqrt{2}}(x_{i} +x_{j} ), \frac{1}{\sqrt{2}} (-x_{i} +x_{j} ))\big) \in L^{2} (\R^{2}).
\end{equation}
In the picture, the dashed line in each beam splitter indicates the input port of the "primary mode", i.e. the mode associated with the first component of the tensor product $L^{2}(\R)\otimes L^{2}(\R)\simeq L^{2}(\R^{2})$ in the description of equation (\ref{split}). The beam splitters are indexed so that the first index indicates the primary mode.

Let $\vert \sqrt{2} z\rangle$ be the coherent input state for mode 4. We detect the scaled number differences $\frac{1}{\vert z\vert} N^{-}_{13}$ and $\frac{1}{\vert z\vert } N^{-}_{24}$, 
where $N^{-}_{ij} =\overline{I_{i} \otimes N_{j} -N_{i}\otimes I_{j}}$, so that the joint detection statistics are described by the unique spectral measure extending the set function 
$$
(X,Y) \mapsto P^{\vert z\vert^{-1} N^{-}_{13}} (X) \otimes P^{\vert z\vert^{-1}N_{24}^-} (Y) = \mathsf D_1(X)\otimes\mathsf D_2(Y),
$$
where the operator acts on the entire four-mode field.

Let $\rho=|\psi\rangle\langle\psi|$ and $\sigma$ be the input states for mode 1 and 2, respectively. Then the state of the four-mode field after the combination of the beam splitters 
and the phase shiter is 
$$
W_{\rho,\sigma,z,\xi}=
U_{13}\otimes U_{24}\left(\, B_{12}(\rho\otimes \sigma) B_{12}^{*} \otimes \vert z\rangle\langle z \vert \otimes \vert ze^{i\xi}\rangle\langle ze^{i\xi} \vert\, \right) U_{13}^{*} \otimes U_{24}^{*}.
$$
We regard $\sigma$, $\vert \sqrt{2}z\rangle $ and $\xi$ as fixed parameters, while $\rho$ is the initial state of the object system, i.e. the signal field.  
The detection statistics then define an observable $G^{z,\sigma,\xi}:\mathcal{B}(\R^{2} )\rightarrow L(\hil_{1})$ on the signal field via
$$
\tr [\rho G^{z,\sigma,\xi}(X\times Y)  ]= \tr [W_{\rho,\sigma,z,\xi} \mathsf D_1(X)\otimes\mathsf D_2(Y)].
$$
This is the signal observable measured  by the detector.

Let $G^{T}$ denote the covariant phase space observable generated by 
a positive trace one operator $T$, that is,
\begin{equation}\label{obs}
G^{T}(Z) =\frac{1}{2\pi} \int_{Z} W_{qp} T W_{qp}^{*} dqdp
\end{equation}
for all $Z\in\mathcal{B}(\R^{2})$, where $W_{qp}$, $(q,p)\in\R^{2}$, are the Weyl operators associated with the position and momentum operators $Q$ and $P$. 
Let $C:\hil_{2}\rightarrow\hil_{1}$ denote the conjugation map, i.e. $(C\varphi )(x) =\overline{\varphi(x)}$ in the coordinate representation, 
and let  $(r_{n})$ be any sequence of positive numbers tending to infinity. It was shown in \cite{JukkaVII} that 
the measured observable
$G^{r_{n}, \sigma,\frac{\pi}{2}}$
approaches with increasing $n$  the phase space observable generated by $C\sigma C^{-1}$, that is, 
$$
\lim_{n\rightarrow \infty} G^{r_{n}, \sigma,\frac{\pi}{2}} (Z) =G^{C\sigma C^{-1}}(Z)
$$
in the weak operator topology, for any $Z\in\mathcal{B}(\R^{2} )$ such that the boudary $\overline Z\cap \overline{Z'}$ has zero Lebesque measure.

In general, it is  difficult to determine the domains of the moment operators of the covariant phase space observable $G^{C\sigma C^{-1}}$. 
However, if the generating operator $C\sigma C^{-1}$ is such that $Q^{k}\sqrt{C\sigma C^{-1}}$ and $P^{k}\sqrt{C\sigma C^{-1}}$ are Hilbert-Schmidt operators for all $k\in\N$, 
then according to \cite[Theorem 4]{JukkaIV} we have
\begin{eqnarray}
G^{C\sigma C^{-1}}_{1}[k] &=& \sum_{n=0}^{k} \binom{k}{n} (-1)^{n} \tr[\sigma Q^{n}_{2}] Q^{k-n}_{1},\label{marginaalit41}\\
G^{C\sigma C^{-1}}_{2}[k] &=& \sum_{n=0}^{k} \binom{k}{n} (-1)^{n} \tr[\sigma P^{n}_{2}] P^{k-n}_{1}.\label{marginaalit42}
\end{eqnarray}

\section{Simultaneous measurements of $\sfq$ and $\sfp$}\label{simultaneous}

In the three different measurement models considered above, the actually measured  observable is a covariant phase space observable $G^T$
for an appropriate generating operator $T$. 
Hence, the marginal observables $G_{1}^T$ and $G_{2}^T$ are  convolutions of
the sharp position and momentum observables with the Fourier related probability densities $f$ and $g$ defined by $T$, 
respectively. Indeed, if $T=\sum_it_i|\eta_i\rangle\langle\eta_i|$ is the spectral decomposition of $T$, then
$f(q)=\sum_it_i|\eta_i(-q)|^2$ and $g(p)=\sum_it_i|\hat{\eta_i}(-p)|^2$.
Due to this structure, the moment operators of the marginal observables $G_{1}^T$ and $G_{2}^T$ can be written in simple forms as polynomials of either $Q$ or $P$. 
That is, for any $\psi\in\hil$, 
\begin{eqnarray*}
\langle \psi \vert  G_{1}^T[k] \psi\rangle &=& \sum^{k}_{i=0} s^{Q}_{ki} \langle\psi\vert Q^{k-i}\psi\rangle,\\
\langle \psi \vert  G_{2}^T[k] \psi\rangle &=& \sum^{k}_{i=0} s^{P}_{ki} \langle\psi  \vert P^{k-i}\psi\rangle,
\end{eqnarray*}
where the coefficents $s^{Q}_{ki}$ and $s^{P}_{ki}$ depend on the model in question and $s^{Q}_{k0} =s^{P}_{k0} =1$ in each case. 
From these, the recursion formulae for the moments of the position and momentum distributions 
$\rho^{\sfq}$ and $\rho^{\sfp}$,
with $\rho=|\psi\rangle\langle\psi|$,  
of the object to be measured can be computed:
 \begin{eqnarray}
\langle \psi \vert Q^{k} \psi\rangle &=& \langle\psi\vert G_{1}^T[k]\psi\rangle - \sum^{k}_{i=1} s^{Q}_{ki} \langle\psi\vert Q^{k-i}\psi\rangle,\label{rekonstruktio1}\\
\langle \psi \vert P^{k} \psi\rangle &=&\langle \psi \vert G_{2}^T[k]\psi\rangle - \sum^{k}_{i=1} s^{P}_{ki} \langle\psi  \vert P^{k-i}\psi\rangle.\label{rekonstruktio2}
\end{eqnarray}
If $\psi$ is chosen to be, for example, a linear combination of Hermite functions, 
the distributions $\rho^{\sfq}$ and $\rho^{\sfp}$
are exponentially bounded and as such, are uniquely determined by their respective moment sequences 
$(\langle\psi \vert Q^{k}\psi\rangle )_{k\in\N}$ and $(\langle\psi \vert P^{k} \psi\rangle) _{k\in\N}$. 
In this sense one is able to measure simultaneously the position and momentum observables $\sfq$ and $\sfp$ in such a vector state
in any of the three single measurement schemes collecting the relevant marginal information. 
Furthermore, since the linear combinations of Hermite functions are dense in $L^{2} (\R )$, their associated distributions $\rho^{\sfq}$ and $\rho^{\sfp}$
suffice to determine the whole position and momentum observables $\sfq$ and $\sfp$ as spectral measures.  

\section{Concluding  remarks}\label{info}

We have shown with three different measurement models that the statistical method of moments allows one to determine with a single measurement scheme  both the
position and momentum distributions $\rho^{\sfq}$ and $\rho^{\sfp}$ from the actually measured statistics for a large class of
initial states $\rho$. 
In each case the actually measured observable is a covariant phase space observable $G^T$ whose generating operator $T$ depends on the
used measurement scheme.
Such an observable is known to be informationally complete  if  the operator $T$
satisfies the condition $\tr [W_{qp}T]\neq 0$ for  almost all $(q,p)\in\R^2$\cite{Ali}. Recently
it has been shown  that this condition is also necessary for the informational completeness of $G^T$ \cite{JW}.
Neither the used  models nor the method of moments depend on this assumption.
Indeed, if, for instance $T=  |\eta\rangle\langle\eta|$,   
with a compactly supported $\eta$, 
so that $G^T$ is informationally incomplete,
the equations (\ref{rekonstruktio1}\,-\,\ref{rekonstruktio2}) can still be used to determine $\rho^\sfq$ and $\rho^\sfp$ provided that these distributions are
exponentially bounded, for instance if $\rho=|\psi\rangle\langle\psi|$, with $\psi$ in  the linear hull of the Hermite functions.
If, however, the phase space observable $G^T$ is informationally complete and if one is able to reconstruct the state $\rho$  from this
informationally complete statistics  $\tr [\rho G^T(Z)], Z\in\hB(\R^2)$, then, of course, one knows the distribution of any observable,
in particular, the position and momentum distributions $\rho^{\sfq}$ and $\rho^{\sfp}$.
However, the reconstruction of the state from such a statistics is typically a highly difficult task, see e.g. \cite{Paris_et_al}. 
In the special case of the generating operator $T$ being the Gaussian (vacuum) state $T=|0\rangle\langle 0|$, the distribution
 $Z\mapsto \tr [\rho G^{|0\rangle}(Z)]$ is the 
 Husimi distribution of the state $\rho$. For that, a reconstruction formula is well known and simple \cite{Dariano1994}.
Indeed, writing $z=\frac 1{\sqrt 2}(q+ip)$, one has $W_{qp}|0\rangle=|z\rangle$, and  $ \tr [\rho G^{|0\rangle}(Z)] = \int_ZQ_\rho(z)\,d^2z$,
with $Q_\rho(z) = \frac 1\pi \langle z|\rho|z\rangle$ being the Husimi  Q-function of the state $\rho$. Using the polar coordinates,  
the matrix elements of $\rho$ with respect to the number basis are
$$
\rho_{n,n+k} =\frac{\sqrt{(n+k)!n!}}{(2n+k)!} \frac{d^{2n+k}f(0)}{dr^{2n+k}},
$$
where 
$$
f(r) =\frac{1}{2}e^{r^2} \int_0^{2\pi} e^{-ik\theta} Q_\rho(re^{i\theta}) d\theta.
$$
It is to be emphasized that the reconstruction of the state requires, however, full statistics of the observable $G^{|0\rangle}$. 
The marginal information, which is used in the method of moments,  is clearly not enough to reconstruct the state even in the case where the position and momentum distributions are exponentially bounded. To illustrate this fact, let us consider the functions $\varphi_{a,b} (q) = \left(\frac{2a}{\pi}\right)^{1/4}  e^{-(a+ib) q^2}$, with $a,b\in\R$, $a>0$.  The Fourier transform of $\varphi_{a,b}$ is
$$
\hat{\varphi}_{a,b}  (p) =\left(\frac{a}{2\pi (a^2 +b^2 )}\right)^{1/4} \exp\left( -\frac{ap^2}{4(a^2 +b^2)}\right) \exp\left( \frac{ibp^2}{4(a^2 +b^2)} -\frac{i}{2} \arctan\frac{b}{a}\right),
$$
and the position and momentum distributions are 
\begin{eqnarray*}
 \vert \varphi_{a,b}(q)\vert^2 &=&\left(\frac{2a}{\pi}\right)^{1/2} e^{-2aq^2}, \\
\vert \hat{\varphi}_{a,b} (p)\vert^2 &=&\left(\frac{a}{2\pi(a^2 +b^2)}\right)^{1/2} e^{-\frac{ap^2}{2(a^2+b^2)}},
\end{eqnarray*}
which are clearly exponentially bounded. For $b\neq 0$, we see that $\rho_1 =\vert \varphi_{a,b}\rangle\langle\varphi_{a,b} \vert$ and $\rho_2 =\vert \varphi_{a,-b}\rangle\langle\varphi_{a,-b} \vert$ are different states, but $\rho_1^Q=\rho_2^Q$ and $\rho_1^P =\rho_2^P$. The marginal probabilities are 
\begin{eqnarray*}
p^{G^{\vert 0\rangle}_1}_{\rho_1}(X) &=\int_X (g\ast \rho_1^Q )(x)dx =\int_X (g\ast \rho_2^Q )(x)dx  &=  p^{G^{\vert 0\rangle}_1}_{\rho_2}(X),  \\
 p^{G^{\vert 0\rangle}_2}_{\rho_1}(Y) &=\int_Y (g\ast \rho_1^P)(y)dy =\int_Y (g\ast \rho_2^P)(y)dy &=  p^{G^{\vert 0\rangle}_2}_{\rho_2}(Y), 
\end{eqnarray*}
for all $X,Y\in\br$, with $g(x) =\frac{1}{\sqrt{\pi}} e^{-x^2}$, so the marginal distributions are equal. It follows that the state cannot be uniquely determined from the marginal information only.

Since the marginal observables $G_1^{T}$ and $G_2^{T}$ are of the convolution form with densities, 
the position and momentum distributions can also be obtained if one is able to  invert the convolution.
Indeed, for any initial state $\rho=|\psi\rangle\langle\psi|$ the marginal distributions $p^{G^T_1}_\rho$ and $p^{G^T_2}_\rho$ have the densities
$f\ast\rho^Q$ and $g\ast\rho^P$,
where
$f(q)=\sum_it_i|\eta_i(-q)|^2$, 
$g(p)=\sum_it_i|\hat{\eta_i}(-p)|^2$,
with 
$T=\sum_i t_i|\eta_i\rangle\langle\eta_i|$, and
$\rho^Q=\vert \psi\vert^2$, 
$\rho^P= \vert \hat{\psi}\vert^2$.
The unknown distributions  $\rho^Q$  and $\rho^P$ can  be solved from the measured distributions $f\ast\rho^Q$
and $g\ast\rho^P$ by using  either the Fourier inversion or the differential inversion method. Like the method of moments,
these methods have their own specific restrictions.
In fact, by  the Fourier theory, one has,
for instance, $\widehat{f\ast\rho^Q}=\sqrt{2\pi}\hat{f}\cdot \widehat{\rho^Q}$, so that
$\widehat{\rho^Q} =(2\pi)^{-1/2}\widehat{f\ast\rho^Q}/\hat{f}$,
provided that $\hat{f}$ is pointwise nonzero. If $\widehat{f\ast\rho^Q}/\hat{f}$ is an $L^1$-function, then  the function
$$
\frac 1{2\pi}\int_{-\infty}^\infty e^{ixt}\widehat{f\ast\rho^Q}(t)/\hat{f}(t)\,dt
$$
coincides with the distribution $\rho^Q$ (almost everywhere). Obviously, this puts strong restrictions on the actually measured distribution
$f\ast\rho^Q$ as well as on the `detector' density $f=f(T)$.
The method of differential inversion is known to be applicable whenever the detector densities $f$ and $g$ have finite moments \cite{Hohlfeld}.
In the special case of  $T=|0\rangle\langle 0|,$ so that $f$ and $g$ are the Gaussian $\frac{1}{\sqrt{\pi}} e^{-x^2}$,
one has
\begin{eqnarray*}
\rho^Q(x) &=& \sum_{k=0}^\infty \frac{(-1/4)^k}{k!} \frac{d^{2k}}{dx^{2k}}(f\ast \rho^Q)(x),   \\
\rho^P(y) &=& \sum_{k=0}^\infty \frac{(-1/4)^k}{k!} \frac{d^{2k}}{dy^{2k}}(g\ast\rho^P)(y),
\end{eqnarray*}
provided that the right hand sides exist \cite{Hohlfeld}, which is a further condition on the initial state $\rho$.

To conclude, the statistical method of moments provides an operationally feasible method to measure with a single measurement scheme both the position
and momentum distributions $\rho^Q$ and $\rho^P$ for a large class of initial  states $\rho$, the relevant condition being the exponential boundedness 
of the involved distributions. This method requires neither the state reconstruction nor inverting convolutions.

\appendix
\section{Proof of lemma \ref{apu3}}
If $P$ is a projection in the range of $M$, then $P$ commutes with any effect $M(Z)$, $Z\in \h B(\R^2)$
(see, for instance, \cite[Th. 1.3.1, p. 91]{Ludwig}). Therefore, the marginals $M_1$ and $M_2$ are mutually commutative,
i.e. $M_1(X)M_2(Y)= M_2(Y)M_1(X)$ for all $X,Y\in \h B(\R)$, and the map $(X,Y)\mapsto M_1(X)M_2(Y)$
is a positive operator bimeasure, and extends uniquely to a semispectral measure $G:\h B(\R^2)\to L(\hil)$,
with $G(X\times Y) =M_1(X)M_2(Y)$ for all $X,Y\in \h B(\R)$ (see, e.g. , Theorem 1.10, p. 24, of \cite{Berg_etal}). Let $X,Y\in \h B(\R)$.
Since $M_1(X)$ and $M_2(Y)$ commute and one of them is a projection, we have $G(X\times Y) =M_1(X)M_2(Y) = M_1(X)\land M_2(Y)$,
the greates lower bound of $M_1(X)$ and $M_2(Y)$,
\cite[Corollary 2.3]{Gudder}. Since also $M(X\times Y)$ is a lower bound for $M_1(X)$ and $M_2(Y)$, we obtain
$M(X\times Y)\leq G(X\times Y)$. It follows that $M(Z)\leq G(Z)$ for any $Z\in \h F$, where $\h F$ is the
algebra of all finite unions of mutually disjoint sets of the form $X\times Y$, $X,Y\in \h B(\R)$.
Denote $\h M = \{Z\in \h B(\R^2) \mid M(Z)\leq G(Z)\}$. Now $\h M$ is a monotone class. [If $(B_n)$ is an increasing
sequence of sets of $\h M$, then for any $\vp\in \hil$, we have
$$\bra \vp |M(\cup_n B_n)\vp\ket -\bra \vp |G(\cup_n B_n)\vp\ket=
\lim_n (\bra \vp|M(B_n)\vp\ket-\bra \vp |G(B_n)\vp\ket) \leq 0
$$
because e.g. $Z\mapsto \bra \vp |M(Z)\vp\ket$ is a positive measure. This shows that $\cup_n B_n\in \h M$.
Similarly, we verify the corresponding statement involving decreasing sequences, and thereby conclude that
$\h M$ is a monotone class.]
Since $\h F\subset \h M$, and $\h F$ is an algebra which generates the $\sigma$-algebra $\h B(\R^2)$,
it follows from the monotone class theorem that $M(Z)\leq G(Z)$ for all $Z\in \h B(\R^2)$.
Let $Z\in \h B(\R^2)$, and let $\vp\in \hil$ be any unit vector.
Since $M_{\vp,\vp}$ and $G_{\vp,\vp}$ are probability measures, we get
$$1-M_{\vp,\vp}(Z)=M_{\vp,\vp}(\R^2\setminus Z)\leq G_{\vp,\vp}(\R^2\setminus Z)=1-G_{\vp,\vp}(Z),$$
implying that $\bra \vp|G(Z)\vp\ket \leq \bra \vp|M(Z)\vp\ket$. Since $\vp$ was arbitrary, this implies
$G(Z)\leq M(Z)$. The proof is complete.
\begin{flushright}
 $\square$
\end{flushright}

\

\

\end{document}